\documentclass[aps,prb,a4paper,twocolumn,amssymb,amsfonts,10pt]{revtex4-1}
\usepackage{mathptmx} 
\usepackage{bm}
\usepackage{graphicx}
\usepackage{multirow}
\usepackage{color}
\usepackage{hyperref}
\hypersetup{
pdfborder={0 0 0}, 
colorlinks=true, 
linkcolor=blue,
citecolor=blue,
urlcolor=blue
}

\newcommand{\subsecspacePre}{} 
\newcommand{\subsecspacePost}{} 

\begin{document}

\title{Kelvin waves from vortex reconnection in superfluid helium at low temperatures}

\author{R. H\"anninen}
\affiliation{Low Temperature Laboratory, Department of Applied Physics, Aalto University, FI-00076 AALTO, Finland}

\date{November 10, 2015} 

\begin{abstract}
We report on the analysis of the root mean square curvature as a function of the numerical resolution for a single reconnection of two quantized vortex rings in superfluid helium. We find a similar scaling relation as reported in the case of decaying thermal counterflow simulations by L. Kondaurova {\it et al.} There the scaling was related to the existence of a Kelvin-wave cascade which was suggested to support the L'vov-Nazarenko spectrum. Here we provide an alternative explanation that does not involve the Kelvin-wave cascade but is due to the sharp cusp generated by a reconnection event in a situation where the maximum curvature is limited by the computational resolution. We also suggest a method for identifying the Kelvin spectrum based on the decay of the rms curvature by mutual friction. Our vortex filament simulation calculations show that the spectrum of Kelvin waves after the reconnection is not simply $n(k) \propto k^{-\eta}$ with constant $\eta$. At large scales the spectrum seems to be close to the Vinen prediction with $\eta$ = 3 but becomes steeper at smaller scales. 
\end{abstract}
%
\maketitle

\section{Introduction}\label{s.intro}

The decay of quantum turbulence in the limit of zero temperature is one of the key topics that
seeks an answer. The Kelvin-wave cascade is one promising way of transferring energy from 
intervortex distances to smaller scales where the classical Kolmogorov cascade is 
impossible\cite{Svistunov1995,KS2004,LN2010,SoninPRB2012}. 
There the weak nonlinear coupling between different scales of helical distortions (=Kelvin waves) 
transfers the energy to ever smaller scales until dissipation may become possible, e.g., via
phonon emission.

Reconnections and interactions between neighboring vortices are expected to work as a drive that 
transfers energy from large scales and feeds the Kelvin-wave cascade. At finite temperatures the 
weak Kelvin cascade is easily suppressed by the mutual friction dissipation\cite{BoueEPL2012,KondaurovaPRB2014}. 
In order to understand the decay processes in the zero temperature limit at scales around and 
smaller than the intervortex distance, we must be able to distinguish between the roles 
of the Kelvin-wave cascade, the Kelvin waves directly induced by the reconnection events, and 
the mutual friction dissipation.

Here we concentrate on the Kelvin waves generated by a single reconnection event and investigate
whether a Kelvin cascade originates from the reconnection event. We also illustrate how
Kelvin waves are damped by mutual friction. Our analysis at zero temperature reveals that 
soon after a reconnection the vortex becomes filled with Kelvin waves of different scale 
and that the presence of the shortest scales is limited only by the resolution. However, 
in accordance with earlier work\cite{HanninenPRB2013},
the evidence for the existence of the Kelvin cascade remains weak since the observations 
can be explained by a redistribution of Kelvin waves that are 
initially ``packed inside the reconnection cusp''. 

Our paper is organized such that in Sec.~\ref{s.model} we review the relation between 
the rms curvature and the Kelvin spectrum. We also take into account the numerical contribution from 
a single sharp cusp. In Sec.~\ref{s.vorKWs} we consider a straight vortex with Kelvin waves and 
show how the decay of rms curvature by mutual friction can be used to identify the Kelvin spectrum. 
We additionally illustrate that the mean curvature can strongly increase without changes in 
the Kelvin spectrum. In Sec.~\ref{s.rectworing} we consider the actual reconnection event and 
use the results from previous sections to interpret these results.

\section{Model}\label{s.model}

In superfluid $^4$He the vorticity appears in the form of quantized vortices, with 
circulation quantum $\kappa = h/m_4 \approx$ 0.0997 mm$^2$/s, whose core size
is of the order of 1 \AA. Ignoring compressibility effects, the vortex dynamics can be 
modeled by using the vortex filament model where the vortices are considered as line 
defects and the superfluid velocity field is given by the Biot-Savart law. The filament 
model should be a good model for $^4$He-II where the core radius is five to six orders of magnitude
below the typical experimental intervortex distance. The model is extensively described, 
e.g., in Ref.~[\onlinecite{schwarz85}]. Numerically the vortex, a three-dimensional curve 
${\bf s}(\xi,t)$, where $\xi$ is the length along the vortex and $t$ is time, is described 
by a sequence of points whose motion is here followed by the fourth order Runge-Kutta method.
Our numerics is described in more detail in Ref.~[\onlinecite{HanninenPRB2013}], 
where we analyzed a similar situation of reconnecting vortex rings but concentrated 
on the dissipation due to mutual friction. 

\subsecspacePre
\subsection{Kelvin spectrum and rms curvature}\label{s.KWCcrms}
\subsecspacePost

The identification of helical Kelvin waves in a vortex tangle is a delicate 
problem\cite{HietalaJLTP2013}. A proper identification is currently possible only in a few 
rare cases. In the case of a straight vortex, taken to be along the $z$ axis, and when the 
perturbations can be presented as $w(z) = x(z)+{\rm i}y(z)$, the Fourier transformation
of $w(z)$ determines the Kelvin-wave amplitudes $w_k$. The Kelvin occupation spectrum 
is typically defined as
\begin{equation}
n(k) = |w_k|^2+|w_{-k}|^2 \, , \, \, k>0 \, .
\end{equation}
One should note that here the $k$ vector is a one-dimensional vector that defines 
the wavelength for the Kelvin waves and should not be confused with the amplitude of the
three-dimensional ${\bf k}$ vector that is typically used when writing the energy (velocity) spectrum.

To provide a rough measure of the Kelvin waves the distribution of curvature radii can
be extracted from a vortex filament calculation. However, in the general case, to derive 
the Kelvin spectrum from the curvature spectrum leads to an insolvable inverse problem.
In case of a straight vortex, and in the limit of small Kelvin amplitudes, these two 
spectra are closely related. Unfortunately, the curvature spectrum is typically much 
too noisy and no proper spectrum can be determined from the curvature data. 
Therefore, the analysis in Ref.~[\onlinecite{KondaurovaPRB2014}] was restricted to
the calculation of the root mean square (rms) curvature, $c_{\rm rms}$:
\begin{equation}
c_{\rm rms}^2 = \frac{1}{L}\int c(\xi)^2d\xi \, ,
\end{equation} 
where $c(\xi)=|{\bf s}''(\xi)|$ is the local curvature at $\xi$, and $L$ is the total vortex length. 
For small Kelvin amplitudes (more precisely when $|w'(z)|=|dw/dz|\ll 1$) and when the energy is taken to 
be proportional to the vortex length [localized induction approximation, (LIA)], the energy spectrum 
$E(k)\propto k^2n(k)$ for Kelvin waves with wave number $k$ can be related to the rms curvature 
by\cite{KondaurovaPRB2014}
\begin{equation}
c_{\rm rms}^2 \approx \frac{4\pi}{\Lambda\kappa^2}\int_{k_{\rm min}}^{k_{\rm max}} k^2E(k) dk \, .
\label{e.crmsEk}
\end{equation} 
The term $\Lambda=\ln(\ell/a_0)$ depends only weakly on the intervortex distance $\ell$ and 
on the core radius $a_0$. Here the Kelvin spectrum is assumed to be valid from $k_{\rm min}$ 
up to some cutoff scale $k_{\rm max}$, above which the spectrum is suppressed (either due
to numerical cutoff or, e.g., by suppression due to mutual friction). 

Since the predictions for the Kelvin-wave spectrum vary from $E(k) \propto 1/k$, 
in the case of strong wave turbulence with the Vinen spectrum\cite{VinenJLTP2002}, 
to $E(k) \propto k^{-5/3}$ with the L'vov-Nazarenko spectrum\cite{LN2010}, 
the rms curvature is not much affected by the scales near $k_{\rm min}$. The dominant 
contribution originates from scales near $k_{\rm max}$ and may therefore be limited 
by the numerical resolution of the calculation. 

The different proposals for Kelvin-wave spectra by Vinen\cite{VinenJLTP2002} (V), 
Kozik-Svistunov\cite{KS2004} (KS), and L'vov-Nazarenko\cite{LN2010} (LN) in terms 
of $c_{\rm rms}$ are asymptotically given by\cite{KondaurovaPRB2014}
\begin{eqnarray}
\label{e.crmskmax}
c_{\rm rms}^{\rm V} &=& \Phi k_{\rm max} \nonumber \\ 
\ell c_{\rm rms}^{\rm KS} &=& \Phi(\ell k_{\rm max})^{4/5} \\
\ell c_{\rm rms}^{\rm LN} &=& \Phi(\ell k_{\rm max})^{2/3}\, . \nonumber
\end{eqnarray}
Generally, when the Kelvin spectrum $n(k)\propto k^{-\eta}$, the rms curvature is given
by $c_{\rm rms}\propto k_{\rm max}^{(5-\eta)/2}$. The intervortex spacing $\ell$ 
limits the smallest possible $k$ values, and the term $\Phi$, given by
\begin{equation}
\Phi = \sqrt{4\pi{E}/\Lambda\kappa^2},\hspace{2mm} E=\int_{k_{\rm min}}^{k_{\rm max}}E(k)dk \, , 
\label{e.PhiDef}
\end{equation}
takes into account the fraction from the total energy $E$ (per unit length, per unit mass) 
related to Kelvin waves. 
Because some numerical factors of the order of unity have been neglected
in the above derivation, one should consider $\Phi$ only as an estimate.  

We expect that at low enough temperatures the Kelvin cascade becomes the dominant
dissipation mechanism, which at scales smaller than the intervortex distance transports
energy towards the core scales where it can be dissipated. For this to be true, at
sufficiently low temperatures the Kelvin-wave cascade causes all our numerically 
traceable scales to be occupied by Kelvin waves, and the rms curvature becomes determined by 
the numerical resolution. By repeating simulations with different resolutions one should 
be able to distinguish between different theories. 

This is precisely what was done in Ref.~[\onlinecite{KondaurovaPRB2014}] for 
decaying thermal counterflow, where the scaling for $c_{\rm rms}$ weakly 
supported the explanation in terms of the Kelvin-wave cascade with L'vov-Nazarenko spectrum. 
The complicated situation was exemplified by the fact that the resolution could only 
be changed by a factor of 2.7, and no firm proof was obtained. 

In the following we consider a simpler case: a single reconnection of two initially linked 
vortices. In this case we are able to change the resolution by a greater amount. However, 
first we consider how the rms curvature is affected when a vortex configuration contains a 
sharp cusp where the curvature may diverge (or at least yield values close to the inverse 
core size).

\subsecspacePre
\subsection{Sharp cusp and its effect on rms curvature}\label{s.cuspcrms}
\subsecspacePost

As soon as one introduces a reconnection in the filament model one creates a 
sharp cusp in the vortex configuration. A simplified analytical model for a cusp where the 
curve has a simple discontinuity in the derivate, being smooth elsewhere, does not 
necessarily change the rms curvature by a large amount. However, numerically the effect 
can be much more dramatic. Without additional smoothing, the maximum curvature at the 
cusp is the inverse resolution, and the region for this curvature peak is of the order 
of the point separation ($\Delta\xi_{\rm res}$). Smoothing does not necessarily 
solve the problem, because it only widens the region and restricts the maximum value. 

If we split the integral into the ``smooth part'', far from the cusp, and into the 
region that is strongly affected by the numerics, one obtains that
\begin{equation}
c_{\rm rms}^2 = \frac{1}{L}\int_{\rm smooth} c(\xi)^2d\xi + \frac{b}{L\Delta\xi_{\rm res}} \, .
\label{e.crmsNum}
\end{equation} 
Here the numerical factor $b$ is of the order of unity. It should depend weakly on 
the numerical details and, e.g., on the initial angle between the vortices before 
the reconnection. In the limit of high resolution the last term becomes dominant, 
and the rms curvature due to numerics becomes
\begin{equation}
c_{\rm rms}^{\rm N} = \sqrt{\frac{b}{L\Delta\xi_{\rm res}}} \propto k_{\rm max}^{1/2} \, ,
\label{e.crmskmaxN}
\end{equation}
where $k_{\rm max} = 2\pi/\Delta\xi_{\rm res}$ is determined by the resolution.

If we consider the analytical model (which is based on earlier numerical work) 
for the vortex shape before the reconnection [\onlinecite{BouePRL2013}],   
we get further support for Eq.~(\ref{e.crmskmaxN}). This estimate for the 
prereconnection vortex shape indicates that the cusp sharpens until the minimum distance 
reaches the core size (after which the filament model is not valid any more) such that 
eventually the region, where the curvature is of the order of the inverse core radius, is 
also of the order of the core diameter. Simulations with the Gross-Pitaevskii model indicate 
that the curvature does not increase much from this limit imposed by the core radius
and that it also has a similar magnitude after the reconnection\cite{ZuccherFP2012}. 
Therefore, Eq.~(\ref{e.crmskmaxN}) should originate already from the prereconnection 
dynamics where $k_{\rm max}$ is given by the core size. 

Even if the cusp contribution to $c_{\rm rms}$ grows slower with $k_{\rm max}$ than the cascade 
predictions in Eqs.~(\ref{e.crmskmax}), it may give the dominant contribution when 
the Kelvin amplitudes are small ($\Phi\ll 1$) and when the resolution is not high enough, 
as shown below. Its contribution also increases with increasing number of reconnections. 

In addition to the rms curvature, we also use the mean curvature $c_{\rm ave}$ in our 
analysis:
\begin{equation}
c_{\rm ave} = \frac{1}{L}\int c(\xi)d\xi \, .
\label{e.cave}
\end{equation} 
Based on similar arguments as used for the rms curvature in Eq.~(\ref{e.crmsNum}), 
the contribution from the cusp region should result in a constant contribution that 
does not increase as the resolution is improved, i.e.,
\begin{equation}
c_{\rm ave} = \frac{1}{L}\int_{\rm smooth} c(\xi)d\xi + \frac{\tilde{b}}{L} \, ,
\label{e.caveNum}
\end{equation}
where $\tilde{b}$ is another constant of the order of unity. Because $c_{\rm ave}$ 
is sensitive to the phases of the Kelvin waves it cannot be directly related to
the Kelvin spectrum.

\section{Kelvin waves on a straight vortex}\label{s.vorKWs}

In order to better understand simulations where Kelvin waves are produced by 
a single reconnection event, we first consider two simplified sets of simulations 
for a straight vortex where the determination of the Kelvin spectrum is less 
challenging. These calculations model the time development {\it after} the reconnection. 
In both cases we occupy the straight vortex, taken to be along the $z$ axis, with a 
known Kelvin spectrum parametrized as
\begin{equation}
n(m) = |w_m|^2+|w_{-m}|^2 = A^2m^{-\eta}\, , \,\, m=1,2,3,\ldots \, .
\label{e.KWspectrum}
\end{equation}
Here $A$ determines the amplitude of the Kelvin spectrum,
and the $w_m$'s can be obtained by Fourier transformation of $w(z)$, provided that 
the configuration remains single valued. At the instant when the
single valuedness breaks, $|dw/dz|$ diverges.
The $k$ vector $k=2\pi{m}/L_z$ is given by the mode number $m$ and period $L_z$.
In all these calculations we have fixed the reactive mutual friction parameter $\alpha'=0$. 

We can make an analytical approximation for the time development of $c_{\rm rms}$ if 
we assume that the Kelvin-wave cascade is absent such that every Kelvin mode decays 
exponentially as $w_m(t) = w_m(0)e^{-t/\tau_{\alpha m}}$, where $\tau_{\alpha{m}}=\tau_{11}/(\alpha{m^2})$.
With $\tau_{11}$ we denote the decay time for mode $m = 1$ when $\alpha = 1$. 
We neglect the weak logarithmic $m$ dependence in $\tau_{\alpha{m}}$. 
By assuming that initially $w_m(0)$ has a spectrum defined by Eq.~(\ref{e.KWspectrum}), we
obtain by using the small amplitude estimation, where $c(\xi)\approx|w''(z)|$ and 
$d\xi\approx dz$, that
\begin{equation}
c_{\rm rms}^2(t) = \frac{(2\pi)^4 A^2}{LL_z^3}\sum_{m>0}m^{4-\eta}e^{-2\alpha{m^2}t/\tau_{11}}\, .
\label{e.crms2decay}
\end{equation}
Here the vortex length $L=L(t)\approx{L_z}$ depends on time only weakly. 
By taking $k_{\rm max}\rightarrow\infty$ and approximating the sum with an integral we
obtain, for $\eta<5$, the asymptotic decay curve for the rms curvature as
\begin{equation}
c_{\rm rms}(t) = \frac{(2\pi)^2A}{L_z\sqrt{2LL_z}}\sqrt{\Gamma{\left(\textstyle\frac{5-\eta}{2}\right)}}
\left(\frac{2\alpha{t}}{\tau_{11}} \right)^{\frac{1}{4}(\eta-5)} \, ,
\label{e.crmsdecay}
\end{equation}
where $\Gamma(a)$ is the gamma function. If for briefness we write 
$c_{\rm rms} = D (\alpha{t})^{-\beta}$, then the decay exponent $\beta = (5-\eta)/4$ 
is simply given by the Kelvin spectrum. Therefore, if the Kelvin-wave cascade is negligible 
during the time window when the decay occurs,
the above law can be used to identify the initial Kelvin spectrum, as shown below.   

If one wants to improve the above approximation, Eq.~(\ref{e.crmsdecay}),
at small  times by taking into account the finite resolution, one should only 
replace the gamma function with the incomplete (lower) gamma function $\gamma(a,x)$ 
as $\Gamma{\left((5-\eta)/2\right)} \rightarrow 
\gamma\left((5-\eta)/2,\, 2\alpha{t}m_{\rm max}^2/\tau_{11}\right)$.
Here $m_{\rm max}$ is the largest Kelvin mode present, which is limited by the 
resolution. At large times Eq.~(\ref{e.crmsdecay}) fails when only a few lowest 
modes are present (dominant). Eventually the decay is simply given by the exponential 
decay of modes $m = \pm{1}$.

\subsecspacePre
\subsection{Decay of rms curvature via mutual friction}\label{s.decayKWbyMF}
\subsecspacePost

\begin{figure}[!t]
\centerline{
\includegraphics[width=0.999\linewidth]{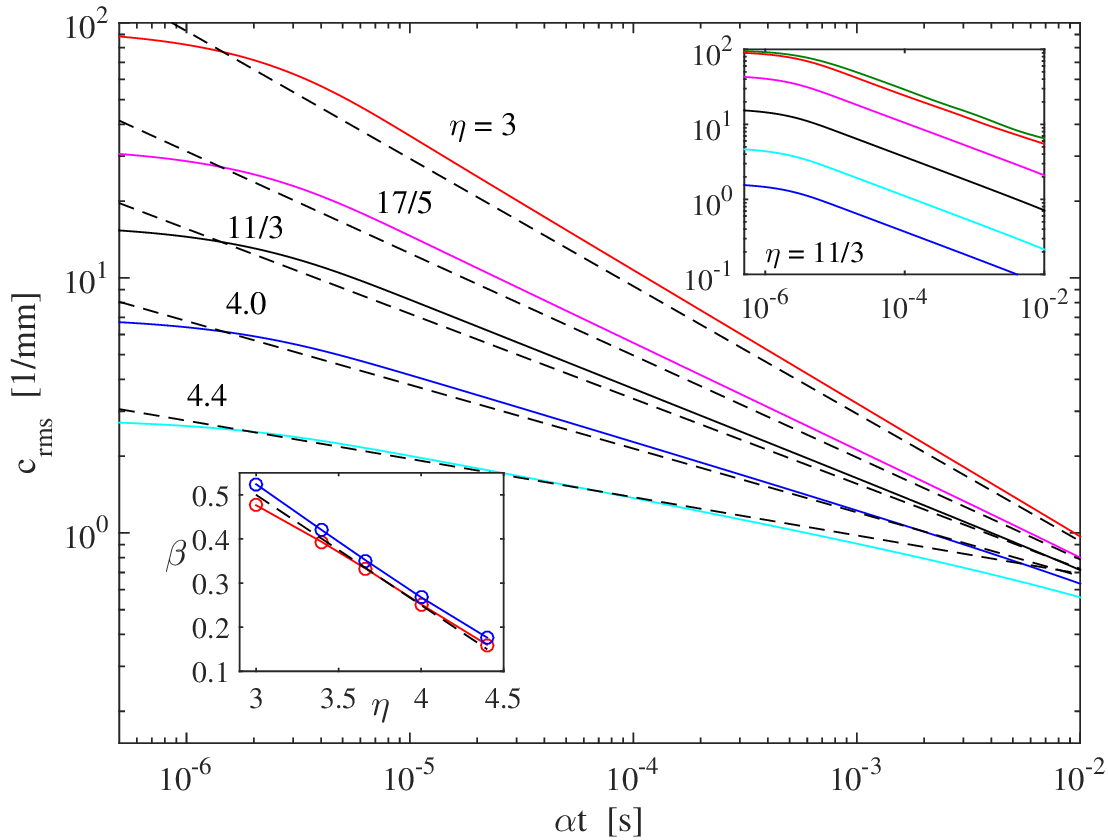}
}
\caption{(Color online)
Decay of $c_{\rm rms}$ for different initial Kelvin spectra on a straight 
vortex with periodic boundary conditions and period $L_z$ = 1 mm. 
The number of points is 1024 and $\alpha = 0.1$. 
In the {\it main panel} the amplitude is $A/L_z=0.01$, see Eq.~(\ref{e.KWspectrum}).
The different spectra are from top to bottom defined by $\eta$ = 3.0 
(red, Vinen), 3.4 (magenta, Kozik-Svistunov), 3.666$\ldots$ (black, L'vov-Nazarenko), 
4.0 (blue), and 4.4 (cyan), respectively. The dashed lines are 
from the analytical model, Eq.~(\ref{e.crmsdecay}).
The {\it upper inset} shows the $c_{\rm rms}$ for the LN spectrum but for different 
amplitudes: from top to bottom $A/L_z$ = 0.3, 0.1, 0.03, 0.01, 0.003, and 0.001, respectively.
The {\it lower inset} relates the decay exponent $\beta$ in 
$c_{\rm rms} \propto t^{-\beta}$ to the exponent $\eta$ in the Kelvin spectrum of Eq.~(\ref{e.KWspectrum}). 
The upper (blue) curve is for small amplitudes with $A/L_z$ = 0.001, and the lower (red)
curve is for rather large amplitudes with $A/L_z$ = 0.1.
The thin dashed line is the exponent appearing in the analytical result [see Eq.~(\ref{e.crmsdecay})].
}
\label{f.decayKWbyMF}
\end{figure}

In our first sample case, we use different Kelvin spectra with different amplitudes,
and observe how the rms curvature decays by the dissipative mutual friction $\alpha(T)$. 
We have additionally assumed that initially the phases of the Kelvin waves are random 
and that $|w_m|=|w_{-m}|$. Figure \ref{f.decayKWbyMF} summarizes our results in case of 
$A = 0.01 L_z$ and $\alpha = 0.1$. The number of points used is 1024 and $L_z$ = 1 mm. 

By repeating the calculations with different values of $\alpha$, we observe that 
the decay curve $c_{\rm rms} = D (\alpha{t})^{-\beta}$ is universal, as predicted by 
Eq.~(\ref{e.crmsdecay}), and the terms $D$ and $\beta$ are not essentially affected by the 
mutual friction parameter $\alpha$. Also, the exponent $\beta$ is almost solely given 
by the Kelvin spectrum exponent $\eta$, as 
shown by the lower inset of Fig.~\ref{f.decayKWbyMF}, where we have used a time window 
$10^{-5}$ s $<\alpha{t}<10^{-3}$ s for the fit. The lower (red) curve is calculated for a 
relatively large amplitude $A/L_z = 0.1$, where 
the energy related to Kelvin waves is of similar order as the energy related to the 
straight vortex. The upper (blue) curve is appropriate for the small amplitude limit 
$A/L_z = 0.001$. 
Further expansion of the amplitude window does not essentially change the results. 

The amplitude $D$ is mainly determined by the Kelvin amplitude $A$. This is illustrated in 
the upper inset of Fig.~\ref{f.decayKWbyMF} using the L'vov-Nazarenko spectrum. The small 
time behavior, which originates from the smallest scales, is sensitive to the resolution
and is consistent with the saturation if the full Eq.~(\ref{e.crms2decay}), or the approximation
described below Eq.~(\ref{e.crmsdecay}), is used.    
Therefore, the power law can be validated also at early times simply by improving the resolution. 

If we use the value $\tau_{11} \approx 0.22$ s from the calculations where only the mode $m=1$ 
is present and approximate that $L\approx L_z$, we obtain nice quantitative agreement with 
Eq.~(\ref{e.crmsdecay}), not only for the exponent, but also for the absolute amplitude. 
This is illustrated in Fig.~\ref{f.decayKWbyMF} with the dashed lines.  
The general tendency for deviations upwards from the theory prediction, Eq.~(\ref{e.crmsdecay}), 
originates from the omitted logarithmic $\ln(m)$ dependence in $\tau_{m\alpha}$.    
The deviations downwards at large times (which is visible in Fig.~\ref{f.decayKWbyMF}
for $\eta$ = 4.4 and 4.0) can be understood to originate from the finite period which 
results in that eventually only the longest Kelvin waves with wavelength equal to $L_z$ 
persist and decay exponentially. Equation~(\ref{e.crms2decay}) models this limit accurately.

The decay curve for $c_{\rm rms}$ is not sensitive to the initial phases of the Kelvin waves, as expected 
from the derivation where the phases are not present. We have also numerically checked that
on the scale of Fig.~\ref{f.decayKWbyMF} one cannot resolve two different decay curves 
with different initial sets of random phases.

\begin{figure}[!t]
\centerline{
\includegraphics[width=0.999\linewidth]{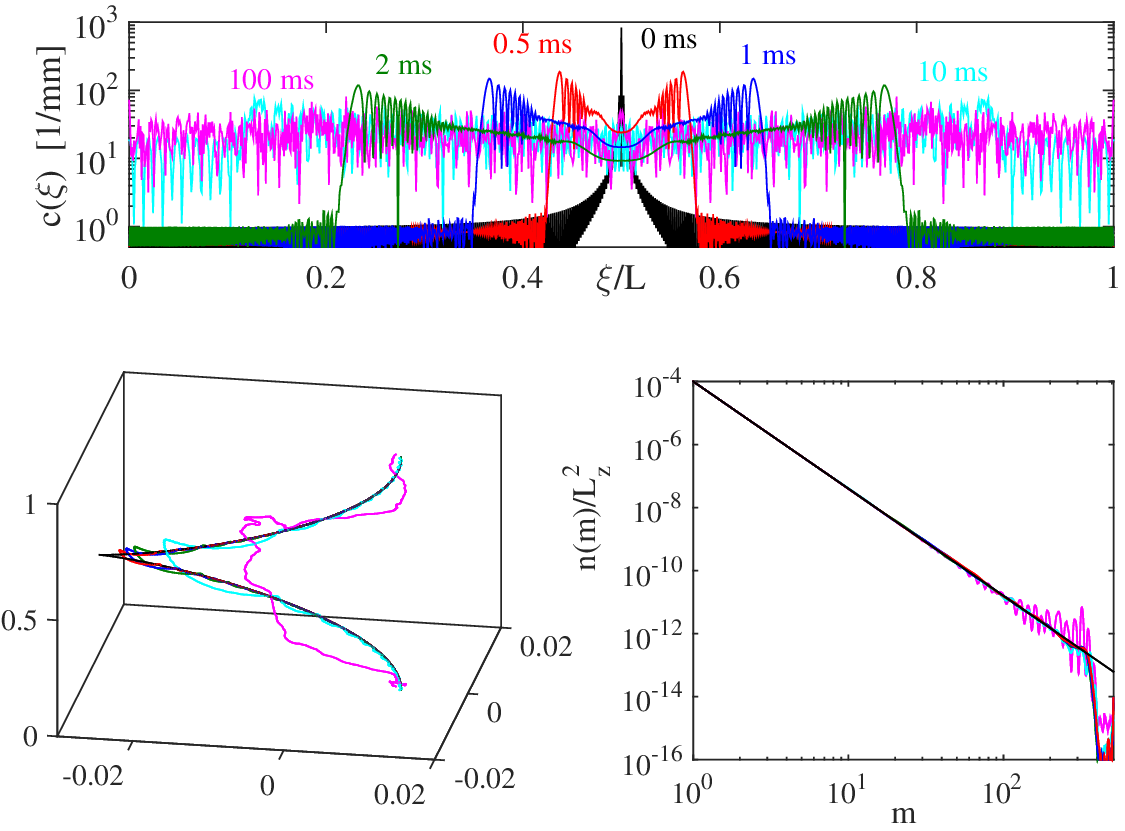} 
}
\caption{(Color online)
Time development of the ``test-cusp'' vortex at $T=0$ that initially hosts the Kozik-Svistunov spectrum. 
The {\it upper panel} illustrates the time development of the local curvature at 
times 0 (black), 0.5 (red), 1 (blue), 2 (green), 10 (cyan), and 100 ms (magenta).
The {\it lower left panel} shows the corresponding vortex configurations,
and the {\it lower right panel} contains the Kelvin spectra where $k=2\pi{m}/L_z$
is given by the mode number $m$.    
}
\label{f.curvtestcusp}
\end{figure}

If we analyze the decay of the different Kelvin modes in Fig.~\ref{f.decayKWbyMF},
we may verify that the different Kelvin modes actually rather well exhibit exponential
decay where the time scale is determined by $\alpha$ and $k$. There are additionally 
fluctuations that are due to interactions between different modes. Also, the decay only 
continues for about four to five orders of magnitude, which is likely due to numerics. We have not
checked the situation for very small $\alpha$ values because it requires a large enough time 
window. This would be important if one seeks the Kelvin cascade, but if the intention is
to determine the initial Kelvin spectrum one is not required to use extremely small
values of mutual friction. Actually, it is much more convenient to use a large enough 
$\alpha$ in order to minimize the effect from the cascade during the decay of $c_{\rm rms}$. 

Therefore, if the characteristic curvatures are much larger than the inverse of the 
intervortex distance $1/\ell$, the decay of $c_{\rm rms}$ should provide a 
potential way to identify the Kelvin spectrum, perhaps also in more complicated tangles.

\subsecspacePre
\subsection{Cusp built by Kelvin waves}\label{s.testcusp}
\subsecspacePost

In our second example case we illustrate how a cusp, where the curvature is strongly peaked, 
can be built on a straight vortex using a Kelvin spectrum defined by Eq.~(\ref{e.KWspectrum}).
We then follow its dynamics both at $T=0$ and $T>0$. 

The phases of the Kelvin waves are typically considered to be random, which results in a wiggly 
looking vortex. However, by specially organizing the phases one may generate a vortex that 
is otherwise smooth but contains a single cusp (per period). 
In Fig.~\ref{f.curvtestcusp} we have built a ``test-cusp'' by setting the same 
constant phase for all the Kelvin waves using the Kozik-Svistunov spectrum. The period  is $L_z$ = 1 mm, 
and the number of points is 1024. All the Kelvin waves have rather small amplitude $A/L_z=0.01$,
but since the phases are the same, $|w'(z)| \approx 0.5$ near the cusp. Therefore, the 
theory predictions made in Sec.~\ref{s.model} are not necessarily valid, e.g.,  
the rms curvature grows slower than the $c_{\rm rms}\propto k_{\rm max}^{4/5}$ prediction. 

\begin{figure}[!t]
\centerline{
\includegraphics[width=0.99\linewidth]{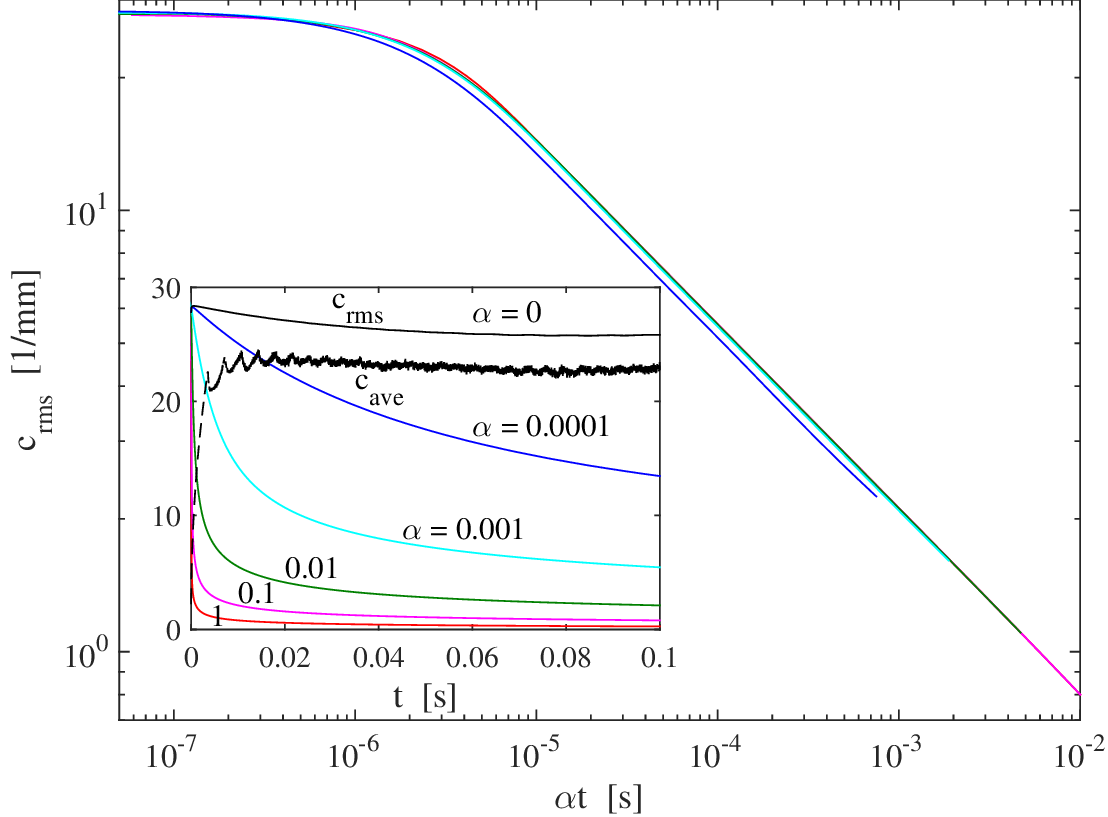}
}
\caption{(Color online)
Decay of the rms curvature when the initial configuration is a specially
arranged sharp cusp with Kozik-Svistunov spectrum, see Fig.~\ref{f.curvtestcusp}. The different 
curves correspond to different values of mutual friction: 
$\alpha$ = 0.0001 (blue), 0.001 (cyan), 0.01 (green), 0.1 (magenta), and 1 (red). 
The {\it inset} shows the time development for $c_{\rm rms}$, and in addition at
$\alpha = 0$ (black) both for $c_{\rm rms}$ (solid) and $c_{\rm ave}$ (dashed), plotted 
on a linear time scale. 
}
\label{f.decaytestcusp}
\end{figure}

The upper panel of Fig.~\ref{f.curvtestcusp} shows the time development of the local curvature
at $T=0$. It seems that with increasing time more small scale structure is developing because
Kelvin waves spread from the cusp and eventually fill the whole vortex. However, if one 
investigates the corresponding Kelvin spectra, shown in the lower right panel of 
Fig.~\ref{f.curvtestcusp}, the amount of Kelvin waves (=$\int_{k_{\rm min}}^{k_{\rm max}}n(k)dk$) 
is not essentially changed because the spectrum is still the same (initial) Kozik-Svistunov spectrum. 
Deformations occur only near the resolution limit, where the numerical errors are largest.

As the cusp relaxes, the phases of the Kelvin waves randomize because all the Kelvin 
waves have different propagation velocity along, and different rotation velocity around, 
the $z$ axis. This results in that the mean curvature, which initially is much smaller 
than the rms curvature, grows and soon has a similar value as $c_{\rm rms}$.
This is illustrated in the inset of Fig.~\ref{f.decaytestcusp}, where we have analyzed 
the decay of the rms curvature by mutual friction, for different values of $\alpha$. 
The main panel shows that the decay follows the universal power law with exponent 
$\beta = 0.4$, derived above for the KS spectrum (Fig.~\ref{f.decayKWbyMF}).

\section{Reconnection of two linked vortex rings}\label{s.rectworing}

\subsecspacePre
\subsection{Numerical setup}\label{s.numerics}
\subsecspacePost

We consider two vortex rings with radius $R$ = 1 mm, initially linked and oriented 
perpendicular to each other. The initial distance between the two vortices is $0.1R$, 
causing that at $T = 0$ they will reconnect approximately at 1.696 s. For dynamics,
see Ref.~[\onlinecite{HanninenPRB2013}].  
Our reconnection method is rather standard: We reconnect 
the vortex segments when the minimum distance becomes smaller than $0.8\Delta\xi_{\rm min}$, 
where $\Delta\xi_{\rm min}$ is the minimum point separation tolerated. The maximum point
separation allowed is $\Delta\xi_{\rm max} = 2\Delta\xi_{\rm min}$. Additionally,
we require that the vortex length must decrease, which approximately means that
the reconnection event is dissipative. 

The time resolution is chosen such that our numerical time step is kept smaller than 
the time scale related to the smallest resolvable Kelvin waves. This implies that our time step 
$\Delta{t} \propto (\Delta\xi_{\rm min})^2$, resulting in that the better the spatial resolution, 
the more difficult it becomes to cover the same overall time window as in the lowest resolution 
run. Another numerical challenge is caused by the fact that 
the computational work per one time step grows as $N^2\propto (1/\Delta\xi_{\rm min})^2$, 
where $N$ is the number of points used to describe the vortex.

\subsecspacePre
\subsection{Finite mutual friction}\label{s.rectworingMF}
\subsecspacePost

\begin{figure}[!t]
\centerline{
\includegraphics[width=0.999\linewidth]{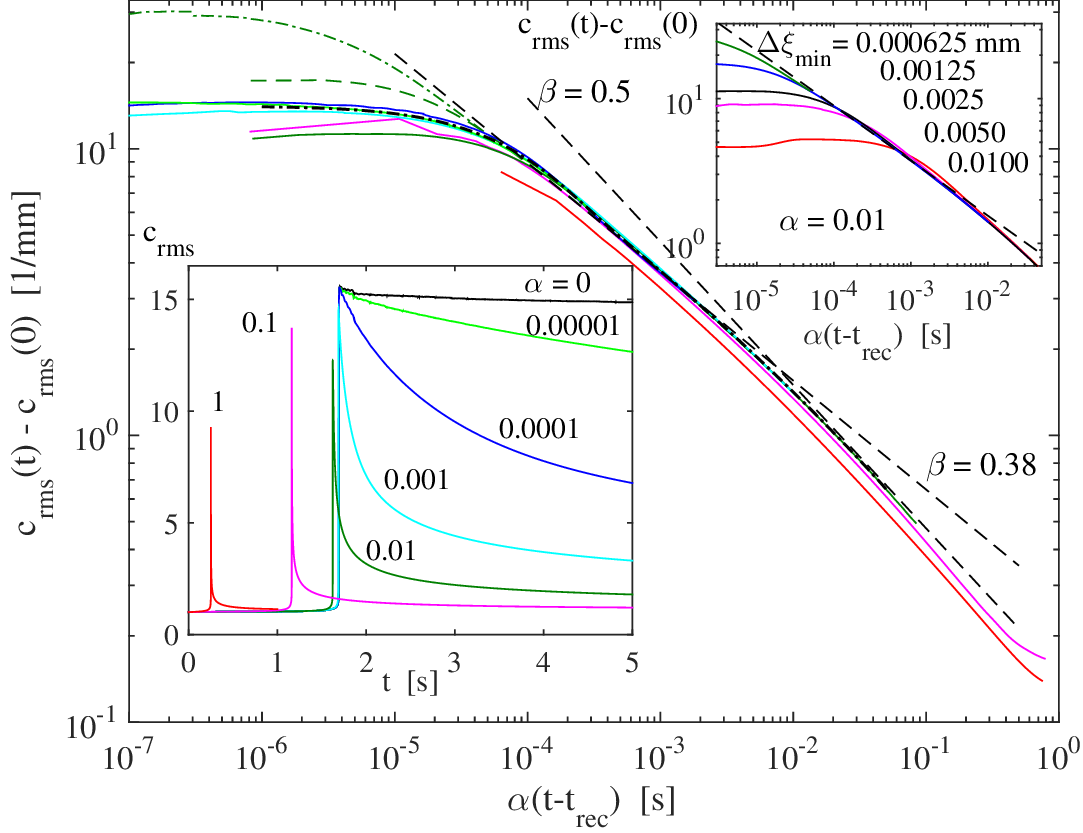}
}
\caption{(Color online)
Time development of the rms curvature for different values of the mutual friction 
parameter $\alpha$ (and $\alpha'=0$).
In the {\it main figure} time is measured from the reconnection instant and scaled by $\alpha$, 
and on the vertical axis the initial curvature $c_{\rm rms}(0)=1/R$ is subtracted from the 
data. Different colors are used for $\alpha$ = 1 (red), 0.1 (magenta), 
0.01 (green), 0.001 (cyan), 0.0001 (blue), and 0.00001 (light green), respectively.
The corresponding maximum times simulated are 1, 9, 10, 12, 20, and 30 s, respectively. 
The black dash-dotted curve is for $\alpha$ = 0.01 where the initial configuration
is taken from $T=0$ simulations, 1 s after the reconnection, and the time (which covers
a time window of 5 s) is measured from this instant.
The curves are calculated with a resolution of $\Delta\xi_{\rm min}$ = 0.0025\,mm, while for 
$\alpha$ = 0.01 results with $\Delta\xi_{\rm min}$ = 0.00125\,mm (dashed, green) and 0.000625\,mm 
(dash-dotted, green)
are also shown. The power-law behavior is illustrated by plotting the $c_{\rm rms}\propto (\alpha{t})^{-\beta}$ 
asymptotics with $\beta$ = 0.5 and 0.38 (dashed) that correspond to $\eta$ = 3 and 3.48, respectively. 
In the {\it lower inset} linear scales are used without any scaling. The color coding is the same as 
in the main figure. The case $\alpha = 0$ is additionally shown by the black curve. 
The {\it upper inset} is for $\alpha$ = 0.01 with five different resolutions indicated on the plot,
together with the $\beta$ = 0.38 asymptotic (dashed line).
}
\label{f.crms4res2}
\end{figure}

Before concentrating on the zero temperature limit we show how finite mutual 
friction affects the vortex dynamics and how it damps the Kelvin waves generated by the 
reconnection cusp. In Fig.~\ref{f.crms4res2} we have plotted the rms curvature 
for one particular resolution, defined by $\Delta\xi_{\rm min}$ = 0.0025 mm, for seven
different values of the mutual friction parameter $\alpha$ (in all cases $\alpha' = 0$).

For $\alpha \sim 1$ the instant of reconnection depends essentially on $\alpha$ 
because the vortex rings shrink more and therefore they also move faster. The
relaxation rate for $c_{\rm rms}$ is strongly affected by the mutual friction. 
However, one may notice that the peak value, just after the reconnection, is 
not substantially affected by $\alpha$. The changes are mainly due to the numerical 
algorithm used for the reconnection. This causes that the distance between the 
neighboring vortices is not exactly the same when the reconnection is made, and 
therefore the sharpness/angle of the cusp may slightly vary.  

If we compare these results with the model system, Sec.~\ref{s.decayKWbyMF}, we
may conclude that the results are consistent with the idea that the amplitude of 
the Kelvin wave decays exponentially with time scale $\tau \propto 1/(\alpha{k^2})$.
These Kelvin waves are generated by the reconnection, and the $\alpha$ scaling shown 
in the main panel of Fig.~\ref{f.crms4res2} is {\it consistent with the idea that 
the Kelvin-wave cascade is absent}. If, at low enough temperatures, the Kelvin cascade 
would become important (compared with mutual friction), then the same scaling would not 
work anymore because the cascade should continuously generate new small scale Kelvin 
waves and $c_{\rm rms}$ should decay slower (or perhaps even grow). 
Note that a similar scaling relation is obtained and plotted for the mutual friction 
dissipation power in Fig.~5 of Ref.~[\onlinecite{HanninenPRB2013}].

The absence of the Kelvin cascade is additionally supported by the decay curve, 
plotted with the black dash-dotted line in Fig.~\ref{f.crms4res2}. Here the decay 
is calculated using $\alpha$ = 0.01, but the initial configuration is taken from the 
zero temperature calculations with $\alpha$ = 0 one second after the reconnection. Since the
decay follows exactly the same curve as the standard $\alpha$ = 0.01 case, the two spectra
are likely the same. Therefore, even during that one second period, which corresponds 
to the Kelvin period with a wavelength of the order of $R$ = 1 mm, the interactions between 
different Kelvin modes have not altered the spectrum which was directly generated 
by the reconnection at $T$ = 0. 

The observed decay exponent in our reconnection calculation (see Fig.~\ref{f.crms4res2}) 
at late times (large scales) is close to $\beta = 0.5$, which corresponds
to the Vinen spectrum ($\eta$ = 3) for the Kelvin waves. This has been suggested, e.g., in 
Ref.~[\onlinecite{VinenJLTP2002}]. At early times, which correspond to 
small scales, the decay slope is slower, and one obtains a reasonable fit
using $\beta$ = 0.38. Using the analytical model in Eq.~(\ref{e.crmsdecay}),
this corresponds to $\eta$ = 3.48. In the upper inset the decay behavior with 
$\alpha$ = 0.01 is illustrated using five different resolutions. The two highest 
resolutions are also included in the main panel of Fig.~\ref{f.crms4res2}. 
The slope at early times is close to the predictions by Kozik-Svistunov\cite{KS2004} 
with $\eta$ = 3.4, but here the spectrum is generated directly by the reconnection,
without the need for a cascade. 

The difficulty in finding a good fit with fixed $\beta$, that would cover the
whole time window for the decay of $c_{\rm rms}$, may indicate that the 
Kelvin spectrum is not necessarily given by Eq.~(\ref{e.KWspectrum}) with 
constant $\eta$. It seems that the smaller the scale the bigger is $\eta$.
This is consistent with the idea suggested by Nazarenko in Ref.~[\onlinecite{Nazarenko2006}],
where he argued that the reconnection drives not only length scales near the intervortex 
distance but also smaller scales. Therefore, the spectrum excited by the reconnection 
would depend on scales. Now these simulations illustrate that the spectrum is not 
simply $n(k)\propto k^{-4}$ that can be associated with a sharp bend.

\subsecspacePre
\subsection{Zero temperature results}\label{s.rectworingT0}
\subsecspacePost

\begin{figure}[!t]
\centerline{
\includegraphics[width=0.99\linewidth]{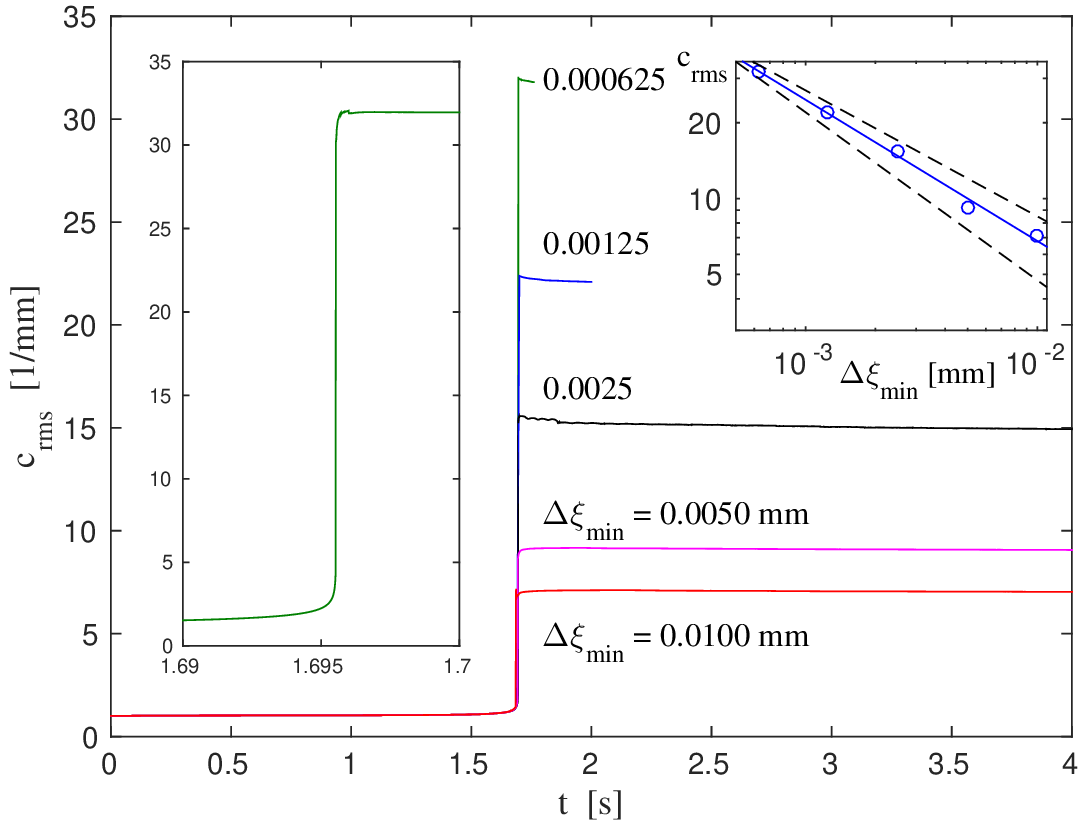}
}
\caption{(Color online)
Time development of the rms curvature at $T=0$ for five different resolutions where the point
separation is kept between $\Delta\xi_{\rm min}\ldots 2\Delta\xi_{\rm min}$. For the two 
highest resolution runs, times well before the reconnection were omitted and the 
calculations were continued using the configuration from the run with one step lower 
resolution where the configuration was still smooth.
The {\it left inset} shows the detailed behavior near the reconnection event when
$\Delta\xi_{\rm min}$ = 0.000625 mm. 
The {\it right inset} shows the scaling relation for the root mean square curvature after 
the reconnection as a function of resolution plus the fit $c_{\rm rms}\propto\Delta\xi_{\rm min}^{-0.56}$ 
(solid line) and the predictions from the L'vov-Nazarenko spectrum (lower dashed) and from the single 
cusp model (upper dashed). 
}
\label{f.crms}
\end{figure}

In the zero temperature limit, where the mutual friction decay time for the single 
Kelvin wave diverges, $c_{\rm rms}$ does not essentially decay after the reconnection kink
has appeared but remains sensitive to the numerical resolution. Figure~\ref{f.crms} 
illustrates the time development of the calculated rms curvature when the resolution 
is changed by a factor of 16 while $\alpha = 0$. The inset illustrates that the curvature 
starts to increase already before the reconnection due to nonlocal interactions. 
This tends to strongly deform the vortex shape near the reconnection point. The noticeable 
jump at the instant of reconnection appears due to the appearance of the sharp cusp which 
is sharpest immediately after the numerical reconnection process. Therefore, 
very locally the curvatures reach the values determined by the resolution. 

After the reconnection the rms curvature may slightly increase and later decrease,
but very soon after the reconnection $c_{\rm rms}$ remains constant. This is 
consistent with the LIA approximation where $c_{\rm rms}$ is one of the several constants 
of motion and where cascade formation is not allowed. We have continued our 
$\Delta\xi_{\rm min}$ = 0.005 mm calculation up to 100 s, and in addition to the small $\sim$1\% 
decay of $c_{\rm rms}$ during this time; it possesses oscillations with amplitude $\sim$1\% and 
period $\sim{20}$ s. This makes it difficult to say accurately from shorter simulation runs 
whether the curvature actually decays, or whether the initial small decay is just part of 
these oscillations. 

If we plot the approximately constant value of $c_{\rm rms}$ after the reconnection as a 
function of the resolution one may observe a scaling relation, similar to the predictions 
above. This is shown in the inset of Fig.~\ref{f.crms}. The fitted exponent is 0.56, which 
is closer to the prediction for a simple cusp than the one originating from the Kelvin 
cascade with L'vov-Nazarenko spectrum. If we want to relate the exponent to a particular 
spectrum, the value corresponds to $\eta=3.88$. Within the cusp model, if we use the minimum 
point separation instead of $\Delta\xi_{\rm res}$ in Eq.~(\ref{e.crmskmaxN}), we obtain 
$b\approx 7$. This is consistent with our prediction (note that our configuration has two cusps).  

\begin{figure}[!t]
\centerline{
\includegraphics[width=0.999\linewidth]{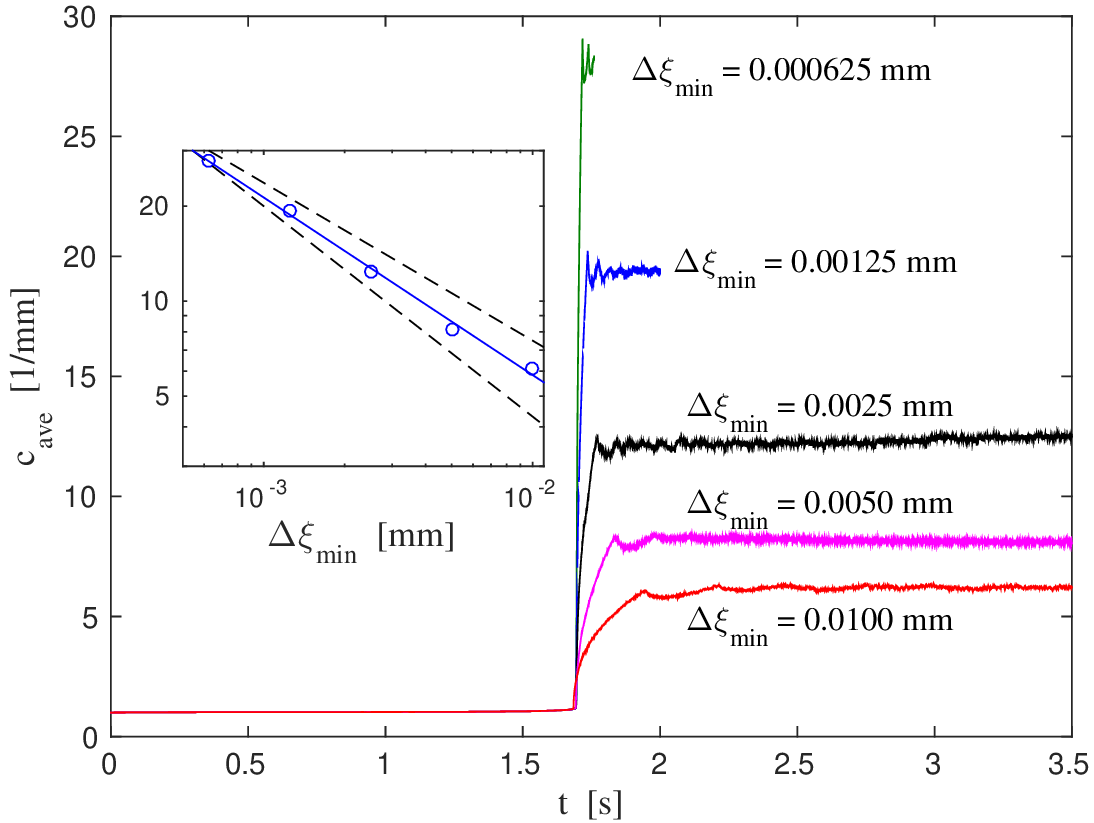}
}
\caption{(Color online)
Time development of the mean curvature for the same five different resolutions which
were used to plot the rms curvature in Fig.~\ref{f.crms}. The {\it inset} shows a similar
scaling relation as that for $c_{\rm rms}$. 
The solid line is a fit, $c_{\rm rms}\propto\Delta\xi_{\rm min}^{-0.56}$, to the data points, 
and the dashed lines are again the predictions from the L'vov-Nazarenko spectrum (lower dashed) 
and from the single cusp model (upper dashed).  
}
\label{f.cave}
\end{figure}

In Fig.~\ref{f.cave} we have plotted the time development of the mean curvature.
One may observe that the large sudden jump, present for the rms curvature, is now missing 
at the instant of reconnection. Rather, the mean curvature starts to increase after the 
reconnection and finally stabilizes to a value very similar to $c_{\rm rms}$. Therefore, 
also the $c_{\rm ave}$ satisfies a similar scaling relation as $c_{\rm rms}$. 

\begin{figure}[!t]
\centerline{
\includegraphics[width=0.99\linewidth]{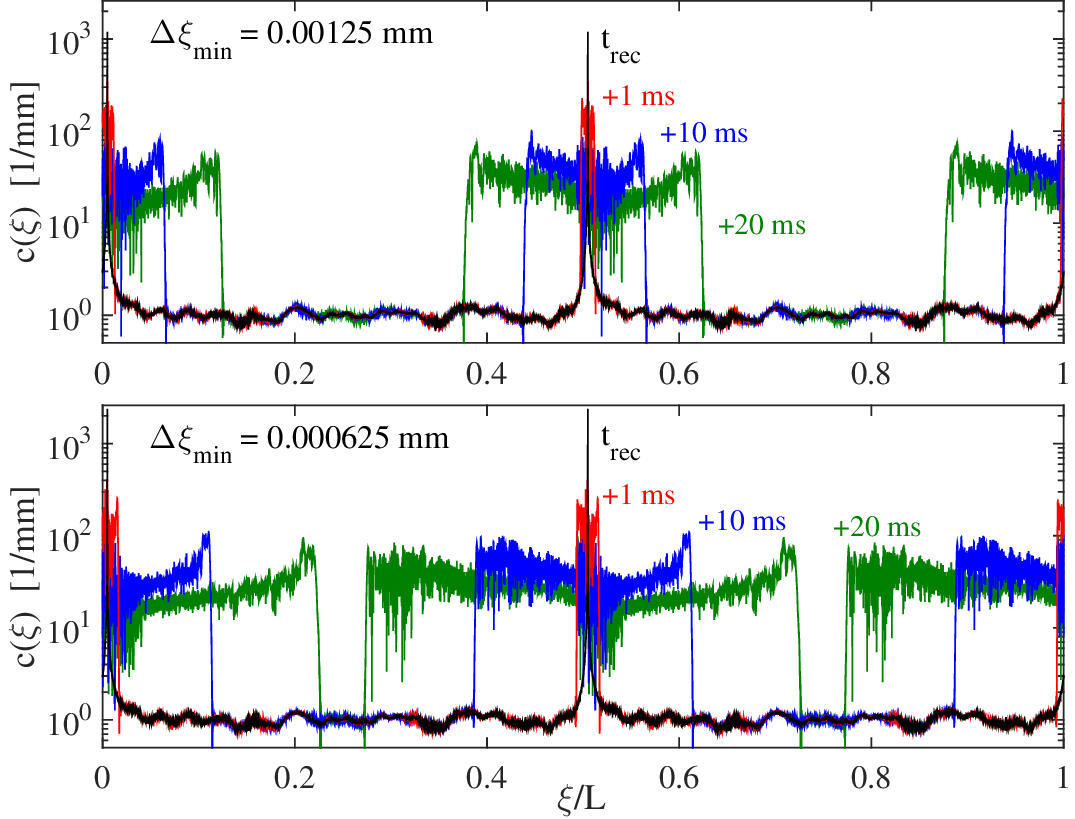}
}
\caption{(Color online)
Time development of the local curvature at $T = 0$ along the vortex after the
reconnection of two rings. The reconnection results in two strong peaks at locations 
$\xi\approx 0$ and $\xi\approx 0.5L$. 
The {\it upper panel} is for the resolution $\Delta\xi_{\rm min}$ = 0.00125 mm, and 
the {\it lower panel} is for twice better resolution, $\Delta\xi_{\rm min}$ = 0.000625 mm. 
The black curve is immediately after the reconnection, and other curves are 1 ms (red), 
10 ms (blue), and 20 ms (green) after the reconnection. 
Note the logarithmic vertical axis and that the maximum speed of the spreading 
Kelvin waves is limited/defined by the resolution. 
}
\label{f.curvatures}
\end{figure}

The time scale for reaching the final value in Fig.~\ref{f.cave} is determined by the 
resolution, and the increase in $c_{\rm ave}$ appears because small scale Kelvin waves travel 
from the reconnection site and quickly fill the whole vortex. The better the resolution, 
the faster the smallest Kelvin waves travel, and therefore the time scale for reaching 
the ``steady-state'' value for $c_{\rm ave}$ is directly proportional to the point 
separation used. This redistribution of the Kelvin waves is illustrated in 
Fig.~\ref{f.curvatures}, where the local curvatures are plotted at different times.

One might argue that the increase in $c_{\rm ave}$ proves the existence of the 
Kelvin cascade because it seems that more small scale structure appears. This is 
possible, but it is not the only possible explanation. First, scaling relations 
similar to those in Eq.~(\ref{e.crmskmax}) cannot be easily derived for $c_{\rm ave}$ from
the Kelvin-wave spectrum. Second, the mean curvature can be strongly increased by 
simply adjusting the phases of the Kelvin waves without changing the Kelvin spectrum,
i.e., without changing the amplitudes of small scale Kelvin waves. 
This is illustrated by our ``test-cusp'' in Sec.~\ref{s.testcusp}, see Fig.~\ref{f.decaytestcusp}.
In the small amplitude limit, which is not necessarily a proper assumption for a cusp, 
the phases of Kelvin waves do not change $c_{\rm rms}$, as shown by Eq.~(\ref{e.crmsEk}). 
Instead, the increase in the average curvature $c_{\rm ave}$ in Fig.~\ref{f.cave} after 
the reconnection appears as the phases ``randomize'' from their initial values at the 
instant of reconnection. 

The above phase randomization can be understood by considering a vortex with a shape 
of a triangle wave whose mean curvature is zero. It has a Fourier representation 
with only odd indices $k=1,3,5,\ldots$ whose amplitudes go like $w_k\propto 1/k^2$, 
corresponding to the Kelvin spectrum $E(k)\propto 1/k^2$
(equal to the Kelvin occupation spectrum $n_k \propto |w_k|^2 \propto 1/k^4$), 
not far from the LN spectrum. The Kelvin waves present in this kind of configuration have
a different frequency around the ``average vortex direction'' (the direction which is realized
when the amplitude of the triangle wave goes to zero) and different propagation 
velocities. Therefore, the vortex has soon a rather different shape with a mean curvature 
much larger than initially. This does not require any cascade. 

One may still wonder why the Kelvin spectrum, determined from the decay curve at small scales
where $\eta \approx 3.5$, does not show up in the scaling relation for $c_{\rm rms}$, which is
close to the cusp model (or alternatively would correspond to Kelvin spectrum with $\eta \approx 3.9$). 
One possible reason for this difference might be that the amplitude of the Kelvin spectrum is too small, 
and therefore the cusp contribution still dominates when calculating the rms curvature. If we 
compare Eqs.~(\ref{e.crmskmax}) and (\ref{e.crmskmaxN}) we find, by setting 
$k_{\rm max}=\pi/\Delta\xi_{\rm min}$, that these estimates give the same value for $c_{\rm rms}$ when
\begin{equation}
\Phi = \pi^{-\gamma}\sqrt{b/L}\,\ell^{1-\gamma}\Delta\xi_{\rm min}^{\gamma-1/2}\, ,
\end{equation}
where $\gamma=(5-\eta)/2$ is the general exponent in Eq.~(\ref{e.crmskmax}) for $k_{\rm max}$.
We may crudely approximate that the energy associated to Kelvin waves is related to the 
increase in vortex length, which is around 1\% in our example case\cite{Note1}, and therefore $\Phi\approx 0.1$.
If we use $b=7$ from our fit, $L\approx 4\pi R$, and additionally approximate that $\ell$ = $R$ = 1 mm,
we obtain that the small scale spectrum with $\eta \approx 3.5$, which dominates the calculation
of rms curvature, should be visible in the scaling relation for $c_{\rm rms}$ if the resolution is
better than $\Delta\xi_{\rm min}$ = 0.01 mm. Our resolution is almost always better than this. 
Because we only see the steeper spectrum in the $c_{\rm rms}$ scaling (Fig.~\ref{f.crms}), the 
Kelvin spectrum near the resolution might be steeper than $\eta$=3.5, or alternative because 
the assumption $|w'(z)|\ll 1$ is likely not valid, the rms curvature grows slower than predicted by 
Eqs.~(\ref{e.crmskmax}). Also, the above approximation is rather sensitive to the value of 
$\Phi$. By using $\Phi=0.05$ the critical resolution increases to $\Delta\xi_{\rm min}$ = 0.0006 mm,
which is the best resolution used. Now, the cusp contribution should always dominate.

\section{Discussion}\label{s.diss}

In more complicated tangles, where vortices experience a large number of reconnections, the 
rms curvature can remain high only if the average time between successive reconnections is
smaller than the time scale for mutual friction damping. This implies that the temperature 
must be low enough, as observed in simulations depicting decaying counterflow 
turbulence\cite{KondaurovaPRB2014}. By considering only a single reconnection event we
are still far from explaining the dissipation in more complicated tangles where 
reconnections occur continually. The future question might be to explain what 
happens to existing Kelvin waves when a new reconnection occurs. Strong local stretching 
caused by a reconnection is likely to damp the pre-existing Kelvin waves near the 
reconnection site.

Our method of determining the Kelvin spectrum from the decay of $c_{\rm rms}$, i.e., obtaining
$\eta$ from the fitted value of the decay exponent $\beta$, could also be applied to more 
complicated vortex tangles. If one needs to avoid reconnections one could 
apply the method separately for each vortex. One may use rather high values
of mutual friction and simulate only over a short time window to obtain the scaling law. 
Actually, with large enough mutual friction the effects from the cascade are minimized
and one obtains more accurately the Kelvin spectrum which describes the initial state
of the vortex. However, with large $\alpha$ we cannot immediately tell whether the spectrum is due 
to a cascade or due to other means like excitations from reconnections.
If the cascade should become important at low enough temperatures, it should then
result in the decay law of Eq.~(\ref{e.crmsdecay}) not being satisfied anymore. 
This change can be verified by repeating the decay analysis at these lower temperatures. 

\begin{figure}[!t]
\centerline{
\includegraphics[width=0.99\linewidth]{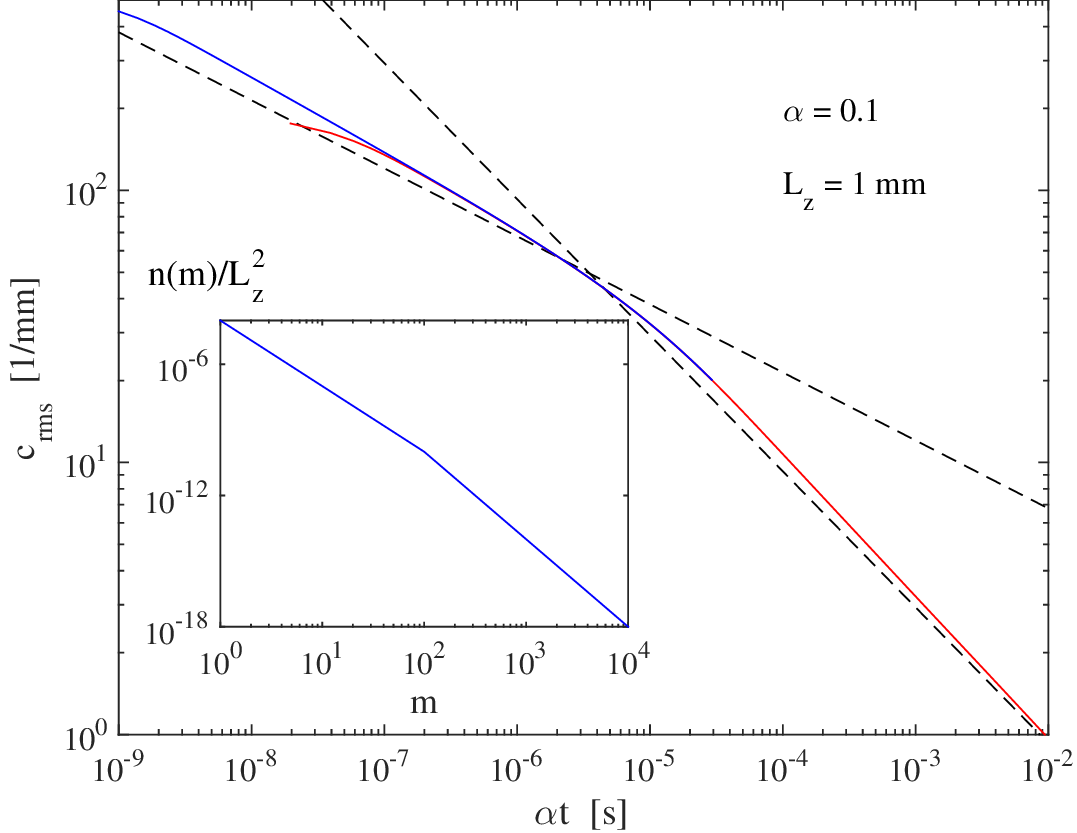}
}
\caption{(Color online)
Decay of the rms curvature in the case of a straight vortex when $\alpha$ = 0.1 and $L_z$ = 1 mm. 
The initial Kelvin spectrum, shown in the inset, is the Vinen spectrum with $\eta = 3$ and $A=0.01L_z$ 
at low $k$'s, while at high $k$'s the spectrum is parametrized by $\eta = 4$ and $A=0.1L_z$. 
The early time behavior (blue curve) is determined using 65536 points on the vortex,
while 8192 points is used to obtain the late time asymptotic (red curve). The dashed 
lines are the analytical results, Eq.~(\ref{e.crmsdecay}), corresponding to both spectra.
}
\label{f.decayEta3and4}
\end{figure}

A nonconstant decay exponent $\beta$ is likely to originate from a Kelvin spectrum 
where $\eta$ is not universal. 
This is supported by the simulations, similar to the ones presented in Sec.~\ref{s.decayKWbyMF}, 
where we have occupied a straight vortex by using the Vinen spectrum at low $k$'s 
($m<100$) with amplitude $A/L_z=0.01$, while using a steeper spectrum with $\eta=4$ 
at high $k$'s with $A/L_z=0.1$, such that the spectrum remains continuous. The resulting 
decay curve has a wide crossover region where $\beta$ changes but both asymptotics 
follow nicely the theory prediction, Eq.~(\ref{e.crmsdecay}). This is illustrated in
Fig.~\ref{f.decayEta3and4}.

The initial approximately constant value of the rms curvature, seen, e.g., at small 
values of $\alpha{t}$ in Fig.~\ref{f.decayKWbyMF}, illustrates that the numerical resolution 
determines when the simulation is numerically in the zero temperature limit. In this
limit even the smallest resolvable scales are not smoothened out by mutual friction. 
From the decay of the smallest scale Kelvin waves one may derive that if 
$\alpha{\Delta{t}}\ll (\Delta\xi_{\rm res})^2/\kappa$ then the effects from mutual 
friction are negligible within times of order $\Delta{t}$. This is consistent
with the observations by Kondaurova {\it et al.}\cite{KondaurovaPRB2014} where 
their simulations at $T\lesssim 0.5$ K are essentially the same as those conducted at 
$T = 0$.

The presence of the Kelvin-wave cascade in simulations has been a controversial topic.
The concern concentrates around the computational resolution required to resolve the 
excitations at short length scales in the zero-temperature 
limit. In Ref.~[\onlinecite{KivotidesPRL2001}] Kivotides {\it et al.} study the reconnection
of four rings in $T$ = 0 simulations and report excitation of Kelvin waves on different
scales, including the appearance of a cascade. The later conclusion is not in agreement
with our findings. However, in the simulations of Ref.~[\onlinecite{KivotidesPRL2001}] 
the extremely crinkled shape of the vortex after the reconnection differs from simulations 
conducted later\cite{HanninenPRB2013,BaggaleyPRB2011} where more effort has been spent 
on the numerical stability and conservation of energy. 
See also Ref.~[\onlinecite{HietalaJLTP2013}] for numerical challenges that appear when 
Kelvin waves are being identified.

\section{Conclusions}\label{s.concl}

Based on vortex filament simulations for a single reconnection event in 
superfluid helium, we have developed an explanation for the scaling relation seen 
for the rms curvature as a function of the numerical resolution. 
In contrast to suggestions by Kondaurova {\it et al.}\cite{KondaurovaPRB2014}, 
our model does not involve the Kelvin cascade but originates from the excitations
created by a reconnection cusp. The distribution of Kelvin waves with different 
wavelengths, which becomes visible after the cusp relaxes, can be understood to arise from the 
prereconnection dynamics where the minimum distance between vortices necessarily sweeps 
all length scales down to the core scale. Therefore, the cusp is built from Kelvin waves 
of different scale, limited only by the numerical resolution, that redistribute when 
the cusp relaxes. 

Thus we conclude that the response to a single reconnection can take place without 
the excitation of a Kelvin-wave cascade. This statement was explicitly tested in 
vortex filament calculations for mutual friction dissipation down to $\alpha \gtrsim 10^{-4}$
and additionally at $T=0$. Instead the reconnection cusp leads to an exponential 
decay of the calculated rms curvature $c_{\rm rms}$ which can be associated with a Kelvin
spectrum $n(k)\propto k^{-\eta}$ in the range $\eta \sim 3\ldots 3.5$. We believe that 
this indicates that a universal Kelvin wave spectrum is not necessarily to be expected
as a response to a single reconnection event. 

\begin{acknowledgments}
This work was supported by the Academy of Finland (Grant No. 218211 and LTQ CoE grant).
I am extremely grateful to M. Krusius for his comments and improvements for the paper. 
I also thank N. Hietala, V. L'vov, and W. Schoepe for their comments and suggestions
and the CSC - IT Center for Science Ltd for the computational resources.
\end{acknowledgments}

\end{document}